\documentstyle[12pt]{article}
\def\lb       {\left( }
\def\rb       {\right) }
\def\lmb      {\left\{ }
\def\rmb      {\right\} }
\def\lbb     {\left[ }
\def\rbb      {\right] }
\def\comma      { \, , }
\def\period     { \, . }
\def\bra#1      { \langle \, #1 \, \vert \, }
\def\ket#1      { \, \vert \, #1 \, \rangle \, }
\def\semiket#1  { \, #1 \, \rangle \, }
\def\del        {  \partial  }
\def\half       {  {1\over 2}  }

\def\abs#1      {  \, \vert #1 \vert \,   }
\def\simg    { \ \raisebox{-.5ex}{$ \stackrel{>}{\sim} $} \ }

\def\vecii#1#2      {  \left(\begin{array}{c}#1\\#2\end{array}\right)  }
\def\veciii#1#2#3   {  \left(\begin{array}{c}#1\\#2\\#3\end{array}\right)  }
\def\matrixii#1#2#3#4            {  \left(\begin{array}{cc}#1&#2\\#3&#4
                                       \end{array}\right) }
\def\matrixiii#1#2#3#4#5#6#7#8#9 {  \left(\begin{array}{ccc}#1&#2&#3\\
                                     #4&#5&#6\\#7&#8&#9\end{array}\right)  }
\def\eqabegin         {  \begin{eqnarray}  }
\def\eqaend           {  \end{eqnarray}  }
\def\nn               {  \nonumber  }
\def\parbigskip        {  \par\bigskip  }

\def\parsmallskip      {  \par\smallskip  }

\def\sectionnumbering { \setcounter{equation}{0}
         \renewcommand{\theequation}{\arabic{section}.\arabic{equation}}}
\def\mysection#1{\addtocounter{section}{1} \setcounter{subsection}{0}
                 \sectionnumbering 
    {\large\bf \par \bigskip \parbigskip \noindent \arabic{section} \quad  
     #1  }   \par \bigskip \noindent}
\def\mysubsection#1{\addtocounter{subsection}{1}
      \par \bigskip \noindent  {\normalsize\bf
      \arabic{section}.\arabic{subsection} \quad #1  } 
   \par \medskip \noindent }
\def\csectionast#1    { \begin{center} \par \bigskip \parsmallskip 
  \noindent  {\large\bf #1  }   \par \bigskip \noindent \end{center} }
\def\kket#1   { \vert #1 \rangle \! \rangle }
\def\bbra#1   { \langle \! \! \langle  #1 \vert }
\begin{document}
%%%%%%%%%%%%%%%%%%%%%
%    cover
%%%%%%%%%%%%%%%%%%%%%
%
%%%%%%%%%%%%%%%%%%%%%%%%%%%%%%%%%%%
\def\papertitlepage{\baselineskip 3.5ex \thispagestyle{empty}}
\def\preprinumber#1#2#3{\hfill \begin{minipage}{4.2cm}  #1
              \par\noindent #2
              \par\noindent #3
             \end{minipage}}
\renewcommand{\thefootnote}{\fnsymbol{footnote}}
%
%%%%%%%%%%%%%%%%%%%%%%%%%%%%%%%%%%%%%%%%%%%%%%%
%
\papertitlepage
\setcounter{page}{0}
\preprinumber{January 1998}{PUPT-1759}{hep-th/9801125}
\baselineskip 1.0cm 
\vspace{2.0cm}
\begin{center}
{\large\bf Propagation of Scalars in Non-extremal Black Hole \\
and Black $p$-brane Geometries}
\end{center}
\vskip 7ex
\begin{center}
     {\sc Yuji ~Satoh
   \footnote[2]{ysatoh@viper.princeton.edu} }\\
 %\vskip -1ex
    {\sl Joseph Henry Laboratories, Princeton University \\
 \vskip -1ex
   Princeton, NJ 08544, USA}
\end{center}
%%%%%%%%%%%%%%%%%%%%%%%
\vskip 10ex
%%%%%%%%%%%%%%%%%%%%%%%
\begin{center} {\large\bf Abstract}
\par \vskip 5ex
%%%%%%%%%%%%%%%%%%%%%%%%%
%%%%%%%%%%%%
\parbox{14cm}{
\baselineskip=3.5ex
We discuss the propagation of scalars in a large class of non-extremal
black hole and black $p$-brane geometries in generic dimensions. We
show that the radial wave equation near the horizon possesses the
$SL(2,R)$ structure in every case; approximately it takes the form of
the wave equation in the $SL(2,R) (AdS_3)$ background and has a
symmetry related to the T-duality of the string model in that
geometry. We see a close connection to two and three dimensional black
holes.  We also find that, in some parameter region, the absorption
cross-sections by the black objects take the form expected from a
conformal field theory.  Our results indicate that some of the
properties known about a certain class of four and five dimensional
black holes hold more generally.
\par 
\vskip 3ex
\noindent
PACS codes: 04.70.Dy, 11.25.Hf \\
%Keywords: \\
}
\end{center} 
\newpage
\renewcommand{\thefootnote}{\arabic{footnote}}
\setcounter{footnote}{0}
\setcounter{section}{0}
\baselineskip = 0.6cm
\pagestyle{plain}
\mysection{Introduction}
The microscopic origin of black hole thermodynamics has been an 
intriguing subject of the quantum theory of gravity. In fact, 
a vast number of works has been devoted for the purpose of understanding 
the Bekenstein-Hawking entropy and the Hawking radiation from a fundamental
theory. In this respect, excellent progress has recently been achieved 
in superstring theory. (For a review see, e.g., \cite{Maldacena}.)
For a certain class of four and five dimensional extremal black holes, 
the entropy formula was derived by counting the degeneracy of the corresponding
Bogomol'ni-Prasad-Sommerfield (BPS) states. It was soon realized that 
such a counting reproduces the black hole entropy for 
slightly non-extremal cases. Furthermore, using a string model 
(the effective string model) based on solitonic solutions called $D$-branes,  
it was shown that the decay rates of some non-BPS string states 
precisely agree with the Hawking radiation from the non-extremal 
black holes \cite{CMDMW}-\cite{KRT}. 

However, the relationship between the non-BPS strings and 
the non-extremal black holes is not completely clear: 
The analysis in string theory is based on $D$-branes at weak coupling
whereas the semi-classical calculation in general relativity  
is valid in the strong coupling regime (see, e.g., \cite{MS}); 
Without the BPS condition,
the extrapolation of the results from weak coupling to strong coupling
is not justified.\footnote{For the arguments regarding this issue, see 
\cite{HSM}.}
We do not have a first-principle derivation of the effective string model
either. Moreover, subsequent detailed analysis showed that, 
in a generic parameter region, one does not find agreement between 
the decay rates from general relativity  and those  
from the simple effective string model \cite{KM}-\cite{KKTR}.

One outstanding feature in these arguments is that the non-extremal 
black holes possess properties suggestive of conformal field theory
(CFT) even beyond the parameter region of the agreement
\cite{Maldacena}, \cite{Larsen}, 
\cite{KM}-\cite{KKTR}, \cite{MS2}-\cite{Hosomichi}.
%\footnote{See, also \cite{early}.}
Let us see this for the 
absorption cross-sections of scalars. First, let us consider a quantum 
mechanical system of a scalar  
whose dynamics is described by some CFT 
(which is not necessarily related to black holes and string theory) 
and suppose that the system is in thermal equilibrium.  
Under this general assumption, the absorption cross-section of the scalar 
is given by \cite{MS2,Gubser}
\eqabegin
  \sigma_{\rm abs} & \sim & 
             \omega^{2l-1} \sinh \lb \frac{1}{2} \beta_H \omega \rb
             \biggm\vert \Gamma \lb h_L + \frac{i}{4\pi} \beta_L \omega \rb 
                         \Gamma \lb h_R + \frac{i}{4\pi} \beta_R \omega \rb 
             \biggm\vert^{2}   
   \comma \label{CFT}
\eqaend  
where $ \omega $ is the frequency;  
$ \beta_L^{-1}(\beta_{R}^{-1})$ is the left-(right-) temperature; 
$ \beta_H/2 = \beta_L + \beta_R $; 
$l$ is the number of derivatives acting on the scalar in the interaction 
term and $h_{L(R)}$ is the left-(right-) dimension of the operator coupled 
to the scalar. Remarkably,  
the absorption cross-sections of scalars by the black holes 
take the same form in some parameter region at low energy.
On the black hole side, $l$ indicates the $l$-th
partial wave, $\beta_H^{-1}$ is the Hawking temperature, $ \beta_{L,R} $
are some geometric quantities associated with horizons and $ h_{L,R} $ are
linear functions of $l$.
The effective string model is a kind of CFT and  
the absorption cross-section takes the form (\ref{CFT}).
Furthermore, one finds the precise agreement including numerical factors
with the  general relativity calculation 
in some parameter region \cite{DM}-\cite{KRT}.  
However, for the semi-classical black holes, 
the expression (\ref{CFT}) is valid in a more general parameter region
which the string model cannot probe, and in which disagreement is found 
\cite{KM}-\cite{KKTR}, \cite{MS2}-\cite{Gibbons}.
This region includes near-extremal Kerr-Newman black holes
in four and five dimensions \cite{MS2,CL} 
which are not supersymmetric even in the extremal limit. 

Another interesting feature in the general relativity calculation
is that the radial wave equations have an $SL(2,R)$ structure
near the horizons;
introducing auxiliary variables, 
they can be rewritten as an eigenvalue equation of the 
Laplace operator on $SL(2,R)$ \cite{CL,Hosomichi}.
In addition, one finds that the wave equations 
have a symmetry similar to T-duality symmetries in string theory 
\cite{CL}. These facts suggest
a relation between the non-extremal black holes and the CFT's
associated with $SL(2,R)$.

Given this state of the problem, one might expect a close 
connection between the non-extremal black holes and CFT's
which are more general than the effective string model. 
Also, it would be interesting
to study how general the above properties of the wave equations and
the absorption cross-sections are. We address this issue in the following.
We will discuss the propagation of scalars in various non-extremal
spherically symmetric black hole and black $p$-brane geometries.  
We will find that the above properties are very general and 
common to all these cases.   

The organization of the paper is as follows. In section 2, we discuss 
the scalar propagation in the Reissner-Nordstrom black hole background
in generic dimensions. 
This provides the simplest case in this paper. 
We discuss it in some detail to make the points 
of the later discussions clear. We find that
the properties discussed above are valid also in this case.
Moreover, we find that the T-duality-like symmetry corresponds 
to the actual T-duality of the string model on $SL(2,R)$. 
We observe a
close relation to the two dimensional $SL(2,R)/U(1)$ black
holes and three dimensional BTZ (B{\~ n}ados-Teitelboim-Zanneli) 
black holes.
We notice that the arguments for the Reissner-Nordstrom black holes
 do not depend on the details of 
the geometry. In section 3, we argue that the properties found in section 2
may hold for
a general class of black objects. We confirm this in the subsequent sections.
In section 4, we discuss the dyonic dilaton black holes by Gibbons and Maeda. 
In section 5 and 6, we discuss charged black holes 
in string theory and a large class of $p$-branes in generic dimensions,
 respectively.
It turns out that the arguments are almost the same in every case.
Summary and discussion are given in section 7.    
\newpage
%%%%%%%%%%%%%%%%%%%%%%%%%%%%%%%%%%%%%%%%%%%%%%
\mysection{Reissner-Nordstrom black holes}
%%%%%%%%%%%%%%%%%%%%%%%%%%%%%%%%%%%%%%%%%%%%%%
%
We begin our discussion with the simplest case, i.e., 
the Reissner-Nordstrom black hole. 
To make the points clear and this article self-contained, 
we discuss this case in some detail.
%
%\newpage
\mysubsection{Space-time geometry}
The metric of the Reissner-Nordstrom black holes in $ D $-dimensional
space-time is given by \cite{TMP}
\eqabegin
  ds_{RN}^2 & = & - f_{RN}(r)  d t^2 + f^{-1}_{RN}(r) 
       d r^2 + r^2 d \Omega_{d+1}^2
  \comma 
\eqaend
where $ d = D - 3 $; $ d \Omega_{d+1}^2 $ is the metric of the unit 
$ (d+1) $-dimensional sphere;
\eqabegin
   f_{RN} (r) &=& \lbb 1 - \lb \frac{r_+}{r} \rb ^d \rbb 
                  \lbb 1 - \lb \frac{r_-}{r} \rb ^d \rbb
   \period   
\eqaend 
The geometry has two regular horizons at $ r = r_\pm $. For $D=4$ and $5$, 
this is a special case of the black holes extensively studied
in string theory \cite{Maldacena} and the extremal geometry becomes 
supersymmetric. Thus, for $D=4$ and $5$, the results in this section 
are included in \cite{HTRDKT,MS2,CL,Gibbons}. 

For later use, we introduce a parameter
$ \mu = r_+^d - r_-^d \equiv r_0^d $ and a new coordinate
$ z = (r^d - r_-^d )/\mu $.
The extremal limit corresponds to 
$ \mu \to 0 $. The outer and the inner horizon are at $ z = 1 $ and $ 0 $ 
respectively. In terms of this coordinate, the metric is written as
\eqabegin
 ds_{RN}^2 & = & -z(z-1) V_{RN}^{-\frac{d}{d+1} } (z) d t^2 + r_0^2 
          \ V_{RN}^{\frac{1}{d+1}} (z)
   \lbb  \frac{1}{d^2} \frac{d z^2}{z(z-1)}  + d \Omega_{d+1}^2 \rbb
  \comma \label{RNV}
\eqaend
where 
\eqabegin
  V_{RN} (z) &=&  \lb \frac{r}{r_0} \rb ^{2(d+1)} 
            \ = \  \lb z + \hat{Q}_+ \rb ^{2(1+\frac{1}{d})}
  \comma 
\eqaend
and $ \hat{Q}_\pm = r_\pm^d/\mu $.

Near the outer horizon, the geometry looks like Rindler space;
\eqabegin
  ds_{RN}^2 & \sim & - \kappa_+^2 \sigma_+^2 d t^2 + d \sigma_+^2 + \cdots
  \period
\eqaend
Here $ \sigma_+ $ is $ (2r_0/d) V_{RN}^{\frac{1}{2(d+1)}}(1) \sqrt{z-1} $ and 
$ \kappa_+^{-2} = (2r_0/d)^2 V_{RN} (1) $. The imaginary time $ it $
is regarded as an angle coordinate in $ (\sigma_+, it) $-plane and its 
period is given by
\eqabegin
  \beta_H^{RN} &=& 
   \frac{2\pi}{\kappa_+} \ = \ \frac{4\pi}{d} r_0 V_{RN}^{\half}(1)
   \ = \ \frac{4 \pi}{d \mu} r_+^{d+1}
  \period
\eqaend
Similarly, near the inner horizon, the metric takes the form
\eqabegin
  ds_{RN}^2 & \sim & - \lb - \kappa_-^2 \sigma_-^2 d t^2 + d \sigma_-^2 \rb
            + \cdots
  \comma
\eqaend
with $ \sigma_- = (2r_0/d) V_{RN}^{\frac{1}{2(d+1)}}(0) \sqrt{z} $ 
and  $ \kappa_-^{-2} = (2r_0/d)^2 V_{RN} (0) $. Then the imaginary time 
is again regarded as an angle coordinate, whose period is  
\eqabegin
  \beta_-^{RN} &=& 
  \frac{2\pi}{\kappa_-} \ = \ \frac{4\pi}{d} r_0 V_{RN}^{\half}(0)
  \ = \  \frac{4\pi}{d \mu} r_-^{d+1}
  \period
\eqaend
\mysubsection{Absorption cross-section of scalars}
Now we consider the propagation of a minimally coupled 
massless scalar in this geometry. Its $ l $-th partial wave 
 satisfies the radial equation,
\eqabegin
    \lbb  d^2 \lmb z (z-1) \del_z \rmb ^2   
    + (r_0 \omega)^2 V_{RN}(z) - z(z-1) \Lambda \rbb \varphi_l &=& 0
    \comma \label{weqRN}
\eqaend  
where $ \omega $ is the frequency and $ \Lambda = l(l+d)$. 
We are interested in the absorption cross-section by the black hole.  
The basic strategy to calculate it is the same as in the literature:
we (i) solve the approximated wave equation in the near and the far region,
(ii) match the two solutions, (iii) compute the fluxes at the 
outer horizon and at infinity and (iv) obtain the absorption cross-section.

Following \cite{MS2}, near the outer horizon
 we approximate $ V_{RN}(z) $ by 
\eqabegin
 V_{RN}(z) \ \sim \ \tilde{V}_{RN}(z) & = &  V_{RN}(1) + (z-1) \del_z V_{RN}(1)
\period 
\label{tildeVRN}
\eqaend 
The validity of the approximation will be 
discussed later in detail. Then the wave equation becomes
\eqabegin
  \lbb  d^2 \lmb z (z-1) \del_z \rmb ^2   
    + (r_0 \omega)^2 \tilde{V}_{RN}(z) - z(z-1) \Lambda \rbb \varphi_l &=& 0
    \period \label{appRN}
\eqaend
From this expression, we find that Eq. (\ref{appRN}) has three regular 
singular points at $ z = 0, 1 $ and $ \infty $. These correspond to the inner
horizon, the outer horizon and infinity, respectively. If a homogeneous linear
ordinary differential equation of the second order has only three regular 
singular points, it is completely characterized by the exponents 
(the roots of the indicial equation) at the regular
singular points. This means that the solutions to (\ref{appRN}) are completely 
determined by the information at the horizons and infinity (in an approximated
sense). We denote the exponents at each singular point by  
$ \nu^{RN(')}_{-} $ for $ z = 0 $, $ \nu^{RN(')}_{+} $ for $ z = 1 $ and 
$ \nu^{RN(')}_{\infty} $ for $ z = \infty $. By the standard procedure, we get
\eqabegin
  \nu^{RN}_+ & = &  - \nu^{RN'}_{+} \ = \ 
    - i \frac{r_0 \omega}{d} \tilde{V}^{\half}_{RN}(1) 
     \ = \ - \frac{i}{4 \pi} \beta_H^{RN} \omega \comma \nn \\
 \nu^{RN}_- & = & - \nu^{RN'}_{-} \ = \
    - i \frac{r_0 \omega}{d} \tilde{V}^{\half}_{RN}(0) 
     \ = \ - \frac{i}{4 \pi} \beta_-^{RN} \Bigl( 1 + {\cal O} 
     (\hat{Q}_+^{-1}) \Bigr) \omega \comma \\
  \nu^{RN}_\infty & = & 1 - \nu^{RN'}_{\infty} \ = \
    \half \lb 1 + \sqrt{1 + 4\Lambda/d^2} \rb \ = \ 1 + \frac{l}{d}
  \nn \period 
\eqaend 
As a check we can confirm Fuchs' relation $ \sum \nu^{RN} = 1$.
Note that $ \nu^{RN(')}_{\pm} $ are essentially the periods of the 
imaginary time, $ \beta_{H,-}^{RN} $. 
Then the set of the solutions is represented by the $ P $-function
 of Riemann,
\eqabegin
  &&
     P \lmb \begin{array}{cccc}
           0 & 1 & \infty &  \\
           \nu^{RN}_{-} & \nu^{RN}_{+} &  \nu^{RN}_{\infty} & z  \\
           \nu^{RN'}_{-} & \nu^{RN'}_{+} & \nu^{RN'}_{\infty } &  
       \end{array} \rmb
    \ = \ z^{\nu^{RN}_{-}} (z \! - \! 1)^{\nu^{RN}_{+}}
     P \lmb \begin{array}{cccc}
           1 & 0 & \infty &  \\
           0 & 0 & a & 1 \! - \! z  \\
           c \! - \! a \! - \! b & 1\! - \! c & b & 
       \end{array} \rmb 
    \comma \nn \\
   &&
\eqaend
where 
\eqabegin
  a &=& \nu^{RN}_{-} + \nu^{RN}_{+} + \nu^{RN}_{\infty}
         \ = \ - \frac{i}{4\pi} \beta_L^{RN} \omega - j_l  \comma \nn \\
  b & = &  \nu^{RN}_{-} + \nu^{RN}_{+} + \nu^{RN'}_{\infty}
         \ = \ - \frac{i}{4\pi} \beta_L^{RN} \omega  + (j_l +1) \comma  \\
  c &=& 1 + 2 \nu^{RN}_{+} \ = \ 1 - \frac{i}{2\pi} \beta_H^{RN} \omega
  \comma \nn
\eqaend
and 
\eqabegin
    \beta_L^{RN} &=& \beta_H^{RN} + \beta_-^{RN} \comma \qquad 
    \beta_R^{RN} \ = \ \beta_H^{RN} - \beta_-^{RN} \comma \nn \\
    j_l  &=& - \nu^{RN}_{\infty} \ = \ - \lb 1 + \frac{l}{d} \rb
   \period
\eqaend
Since the $ P $-function in the right-hand side represents the 
solutions to the hypergeometric equation, the mode which is purely ingoing
at the outer horizon is expressed by the hypergeometric functions as
\eqabegin
  \varphi_l^I &=&  
          z^{\nu^{RN}_{-}} (z-1)^{\nu^{RN}_{+}}  F(a, b, c; 1-z)
   \period 
\eqaend

Using the asymptotics of the hypergeometric functions, we find the asymptotic
behavior of $ \varphi_l^I $ for large $ z $ (i.e., large $r/r_0$),
\eqabegin
 \varphi_l^I & \sim & \frac{\Gamma(c) \Gamma(a-b)}{\Gamma(a) \Gamma(c-b)} 
  \lb \frac{r}{r_0} \rb^{l}
\period
\eqaend
 
In the far region $ r \gg r_+ $ $ (z \gg \hat{Q}_+) $, the wave equation 
simplifies to
\eqabegin
 \lbb \lb r^{d+1} \del_r \rb ^2  + r^{2d} \lb r^2 \omega^2 - \Lambda \rb \rbb
  \varphi_l &=& 0
  \period \label{Bessel}
\eqaend
This is just the wave equation in flat space-time and 
can be solved in terms of the Bessel functions. It turns out that the solution 
which matches to the near-region solution is given by
\eqabegin
  \varphi_l^{II} &=& A \lb \frac{2}{\omega r} \rb ^{d/2} 
          J_{l+d/2}(\omega r)
  \comma
\eqaend 
where $A$ is a normalization constant.
For small $ r \omega $, this behaves as
\eqabegin
  \varphi_l^{II} & \sim & \frac{A}{\Gamma(1+l+d/2)} 
   \lb \frac{\omega r}{2} \rb^{l}
   \period
\eqaend

Matching $\varphi_l^I$ and $\varphi_l^{II}$ in the region 
$\omega^{-1} \gg r \gg r_0 , r_+$, we find that 
\eqabegin
  A &=& \lb \frac{r_0 \omega}{2} \rb^{-l}
   \frac{\Gamma(1+l+d/2) \Gamma(c) \Gamma(a-b)}{\Gamma(a)\Gamma(c-b)} 
\period \label{matching}
\eqaend

To get the absorption cross-section, we use the conserved flux,
\eqabegin
  {\cal F} &=& \frac{1}{2i} \sqrt{-g} g^{rr} 
   \lb \varphi_l^\ast \del_r \varphi_l - {\rm c.c.} \rb
 \comma \label{flux}
\eqaend
where $ g_{\mu\nu} $ and $ g $ are the metric and its determinant, 
respectively.
The absorption probability is the ratio of the ingoing flux
at the outer horizon, ${\cal F}_H$, to the ingoing flux at infinity, 
${\cal F}_{\infty}^{\rm in}$. 
Using the asymptotic forms,
\eqabegin
  \varphi_l^I & \sim &  \ e ^{\nu^{RN}_+ \ln (z-1) } \qquad 
   \qquad \qquad \qquad \qquad \qquad \qquad \qquad \ 
    \mbox{ for } \quad z \to 1  \comma \nn \\
  \varphi_l^{II} & \sim & \frac{A}{\sqrt{\pi}} 
   \lb \frac{2}{\omega r} \rb^{(d+1)/2} 
  \cos \lmb \omega r - \frac{\pi}{4} (2l+d+1)  \pi \rmb
   \qquad \ \! \mbox{ for } \quad \omega r \to \infty
  \comma
\eqaend
we get the absorption probability,
\eqabegin
   P_{\rm abs} &=& 
     \frac{{\cal F}_{H}}{{\cal F}_{\infty}^{\rm in}}
  \ = \ d \mu \beta_{H}^{RN} \abs{ A } ^{-2} \lb \frac{\omega}{2} \rb^{d+1} 
  \period \label{prob}
\eqaend
Finally, 
the absorption probability is converted to the absorption cross-section
by \cite{MG,DGM}
\eqabegin
  \sigma_{\rm abs} &=& 
   \frac{\pi^{d/2}}{d} \lb \frac{2}{\omega} \rb^{d+1}
   \lb l + \frac{d}{2} \rb \Gamma \lb 1+\frac{d}{2} \rb \vecii{l+d-1}{l}
    P_{\rm abs} \period \label{Psigma}
\eqaend
Putting Eqs.(\ref{matching}), (\ref{prob}) and (\ref{Psigma}) together,
we obtain the absorption cross-section,
\eqabegin
  \sigma_{\rm abs} 
 &=& \pi^{\frac{d}{2}} \mu^{1+\frac{2l}{d}} \vecii{l+d-1}{l}
  \frac{(l+d/2) \Gamma(1+d/2)}{ \lbb
    \Gamma(1+l+d/2) \Gamma(1+2l/d) \rbb ^2 }
     \label{abscrossLR} \\
    && \quad \times
   \lb \frac{\omega}{2} \rb^{2l-1}  
    \sinh \lb \frac{1}{2} \beta_H^{RN} \omega \rb
   \biggm\vert \Gamma \Bigl( 1+\frac{l}{d}- \frac{i}{4\pi} \beta_L^{RN} \omega
        \Bigr) 
        \Gamma \Bigl( 1+\frac{l}{d}- \frac{i}{4\pi} \beta_R^{RN} \omega \Bigr)
   \biggm\vert ^2
  \period \nn 
\eqaend
Here we have used the identity 
$ \abs{ \Gamma (1-i\gamma) } ^2 = \pi \gamma/\sinh \pi \gamma $.

For $l=0 $ and $ \omega \to 0 $, $\sigma_{\rm abs}$ reduces to 
the area of the outer horizon as proved in \cite{DGM};\footnote{
The condition of the validity of the matching procedure has been
discussed in \cite{KTr}.}
\eqabegin
  \sigma_{\rm abs} & \to & A_H \period \label{arealaw}
\eqaend 

The above form of the absorption cross-section (\ref{abscrossLR}) 
is the same as Eq.(\ref{CFT}), which is expected from a CFT.  
Therefore, 
if the above expression gives the non-trivial frequency dependence
from the hyperbolic and gamma functions under the condition of validity,  
the Reissner-Nordstrom black holes 
in generic dimensions share the CFT structure 
with the four and five dimensional black holes. 
In the next subsection, we will confirm that this is the case. 
We remark that, for $ D \geq 6 $, we do not have any corresponding 
D-brane configurations near and at extremality. We also note that, 
in the formal extremal limit $r_0 \to 0 $ or low energy limit 
$\omega \to 0$, the solution to Eq.(\ref{weqRN}) is given by the 
hypergeometric functions with $\beta_{L,R}^{RN} = 0 $ without the 
approximation about $V_{RN}$.
%
%
%\newpage
\mysubsection{Validity of the approximation}
We obtained the expression of the absorption cross-section (\ref{abscrossLR}).
Now we discuss the range of validity of 
the approximation. For the time being, we focus on $ \Lambda \neq 0$.

First, let us examine the matching procedure.  In the near region, 
we approximated $ V_{RN} $ by $ \tilde{V}_{RN} $ and used the asymptotic
form for $ z \gg 1 $. In the far region $ r \gg r_+$,
 we used the asymptotic form for 
$ \omega r \ll 1 $. Thus, we need to require the following conditions
so that both approximations are valid at the matching point $ r=r_m$;
\eqabegin
     && 
    (r_0 \omega)^2 \vert V_{RN} - \tilde{V}_{RN} \vert \ \ll \
                   \vert z(z-1) \Lambda - (r_0 \omega)^2 V_{RN} \vert \comma 
     \label{nearcond} 
\eqaend
for $ z \gg 1 $, which means that the error is small enough compared with 
the true total potential term, and 
\eqabegin
    r \gg r_+  \comma && r \omega \ll 1 \period
\eqaend
For $ \Lambda \neq 0 $, these conditions reduce to
\eqabegin
  r_m \gg r_+ \comma && r_m \omega \ll 1 \period \label{validcond}
\eqaend
This indicates that the frequency $ \omega $ should be small enough.

For $\omega \to 0$, Eq. (\ref{abscrossLR}) gives 
$\sigma_{\rm abs} \sim \omega^{2l-1}$. However, we are interested in 
more precise $\omega$-dependence.
To get such a non-trivial $ \omega $-dependence, 
the expression (\ref{abscrossLR}) should be valid in the parameter region
$ \abs{ \nu^{RN}_\pm } \sim \beta_{H,-}^{RN} \omega \sim {\cal O}(1) $.
In fact, $ \beta_H^{RN} \omega $ can be of order unity when 
$ \hat{Q}_+ = r_+^d/\mu $ is sufficiently large.
This is typically achieved near extrimality. 
Notice that we need two scales to achieve this; one scale, say $r_0 \omega $, 
should be very small from the condition (\ref{validcond}),
but we can make another scale $\hat{Q}_+$ very large independently of 
$ r_0 \omega $. This is impossible for geometries with only one scale such as 
the Schwarzschild black holes.  
In turn, for $ \abs{ \nu_+^{RN} } \sim {\cal O}(1) $ and $ \hat{Q}_+ \gg 1 $, 
it follows that $ \beta_H^{RN} \sim \beta_-^{RN} $ and hence 
\eqabegin
  \beta_L^{RN} &=& 2 \beta_H^{RN}  \comma \qquad  \beta_R^{RN} \ = \ 0 
  \period
\eqaend
Thus, the absorption 
cross-section does not take the factorized form into the left and  
right part. Instead, Eq.(\ref{abscrossLR}) should be read as
\eqabegin
  \sigma_{\rm abs} &=&
\pi^{\frac{d}{2}} \mu^{1+\frac{2l}{d}} \vecii{l+d-1}{l}
  \frac{(l+d/2) \Gamma(1+d/2) \Bigl[ \Gamma(1+l/d) \Bigr]^2 }
    { \Bigl[ \Gamma(1+l+d/2) \Gamma(1+2l/d) \Bigr] ^2 }
      \label{abscross} \\ 
    && \quad \qquad \qquad \times
     \lb \frac{\omega}{2} \rb^{2l-1} 
      \sinh \lb \frac{1}{2} \beta_H^{RN} \omega \rb
   \biggm\vert \Gamma \Bigl( 1+\frac{l}{d}- \frac{i}{2\pi}
                 \beta_H^{RN} \omega \Bigr) 
         \biggm\vert ^2
  \period \nn
\eqaend
This form still retains the CFT structure and 
is consistent with the results in \cite{HTRDKT,MS2,CL,Gibbons}.

In the following, we concentrate on the case $ \hat{Q}_+ \gg 1$
and $ \abs{ \nu_\pm^{RN} } \sim {\cal O} (1)$, which we are interested in.
In addition to (\ref{validcond}), 
we need to further check the conditions of validity.
In the far region, it is obvious that the fractional error is at most
$ {\cal O} (r_+/r_m) $. However, we should be careful about 
the error from the near region
other than the matching point, i.e., $r < r_m$.\footnote{
This type of errors was not estimated in some references.
  It seems that we should be careful about 
 the range of validity of those results, in particular, far from extrimality.}
Let us recall that $ \tilde{V}_{RN} $ is the first two terms of the expansion
of $ V_{RN} $ around $ z = 1 $ $ (r=r_+)$. 
Since
\eqabegin
   \del_z V_{RN} &=& \lb 1 + \frac{1}{d} \rb \frac{2 V_{RN}}{z+\hat{Q}_-} 
   \comma
\eqaend
$ \tilde{V}_{RN} $ is a good approximation for 
$ \abs{ z } \ll \hat{Q}_- \sim \hat{Q}_+ $ whereas 
it is not for $ \abs{ z } \sim \hat{Q}_+ $ $ (r-r_+ \sim r_+;\  r \sim 0)$.
In spite of that, 
the approximation can be valid even for $ \abs{ z } \sim \hat{Q}_+ $
if the remaining potential term $ z(z-1)\Lambda $ is large enough.
In fact, we find that the condition (\ref{nearcond}) is satisfied
there if $ r_+ \omega \ll \Lambda $, which is already included 
in (\ref{validcond}). 
 
Therefore, 
the non-trivial $\omega$-dependence in (\ref{abscross}) is actually obtained 
 for
\eqabegin
    r_+ \omega \ll 1 \comma \qquad \hat{Q}_\pm \gg 1 \comma \qquad
    \abs{ \nu_\pm^{RN} } \sim {\cal O}(1)
   \comma
\eqaend 
and the absorption cross-section has the 
CFT structure in this parameter region.
Although the further consideration for $ r < r_m $ 
did not give any additional conditions here, we will see that 
the errors become significant inside the matching point in some cases.

For $ \Lambda = 0 $ $(l=0)$, the term $ z(z-1)\Lambda $ vanishes and 
we need to require
that $ (r_0 \omega)^2 V_{RN},(r_0 \omega)^2 $ $ \times \tilde{V}_{RN} \ll 1 $ 
for $ \abs{ z } \simg \hat{Q}_\pm $. 
This yields just (\ref{arealaw}), which is obtained also from 
(\ref{abscross}) with $ l = 0 $.
 
Let us summarize the discussion.
In order for the approximation to be valid, the frequency $ \omega $ had to
be small enough compared with the size of the black hole. 
In spite of that, the non-trivial $\omega$-dependence 
in (\ref{abscross}) could be obtained for large $\hat{Q}_\pm $. 
As a result, we confirmed that 
the absorption cross-section took the form expected from a CFT.  
The approximation to 
$ V_{RN} $ was valid near the horizons where the other potential term 
$ z(z-1) \Lambda $ was small. Although $ \tilde{V}_{RN} $ was not a good 
approximation far from the horizons including the matching point, 
$ z(z-1) \Lambda $ grew large enough and hence the approximation was valid 
in total for $\Lambda \neq 0$. 
\mysubsection{$SL(2,R)$ structure}
We saw that the absorption cross-section 
takes the CFT structure. Another interesting property found for the black holes
in string theory is that the wave equations have an $SL(2,R)$ symmetry near the
horizon \cite{CL,Hosomichi}. 
In this subsection, we find that this is also the case 
for the Reissner-Nordstrom black holes in generic dimensions.
Moreover, we find a T-duality-like symmetry similar to the one discussed in 
\cite{CL}.
We see that this corresponds to the actual T-duality of the string 
model on $ SL(2,R) $.
We also observe a close relation
to the two dimensional $SL(2,R)/U(1)$ black holes \cite{Witten} and the 
three dimensional BTZ black holes \cite{BTZ}.

We start with a brief review of relevant properties about $SL(2,R)$.
Using analogs of Euler angles, an element $ g_0 \in SL(2,R) $ 
is parametrized as
\eqabegin
  g_0 &=& \ e^{\theta_L \sigma_3/2} 
        \ e^{\rho \sigma_1/2} \ e^{-\theta_R \sigma_3/2}
  \comma
\eqaend
with $ \sigma_i $ the Pauli matrices. In terms of these coordinates, 
the metric of $ SL(2,R) $ takes the form
\eqabegin
  ds^2_{SL} &=& 
    - \lbb \lb \frac{r}{r_0} \rb^2 - 1 \rbb d\theta_+^2 
           +  \lbb \lb \frac{r}{r_0} \rb^2 - 1 \rbb ^{-1} 
             dr^2 + \lb \frac{r}{r_0} \rb^2 d\theta_-^2
  \comma
\eqaend
where 
\eqabegin
  r_0 \theta_{L,R} = \theta_+ \pm \theta_- \comma 
    \qquad r^2 = r_0^2 \cosh^2 (\rho/2) \comma
\eqaend
and we have introduced a dimensionful parameter, $r_0$,
with which the scalar 
curvature is written as $ -6r_0^{-2}$.
The geometry has two Rindler space-like regions;  
the metric becomes 
\eqabegin
   ds^2_{SL} & \sim & \lbb - (\sigma/r_0)^2 d \theta_+^2 + d\sigma^2  \rbb 
   + d \theta_-^2 \comma
\eqaend
for $ r \sim r_0 $ with $ \sigma^2 = r^2 - r_0^2 $ whereas 
\eqabegin
   ds^2_{SL} &\sim & d\theta_+^2 - \lbb dr^2 - (r/r_0)^2 d \theta_-^2  \rbb
  \comma
\eqaend
for $ r \sim 0 $.

The translations along $ \theta_\pm$ are isometries of $ SL(2,R) $. These 
correspond to the vector and axial symmetry of the $SL(2,R)$ WZW model, 
respectively. By gauging either of them, we obtain the Lorentzian 
two dimensional black holes \cite{Witten}. 
The orbifolds with respect to the discrete
symmetries generated by their linear combinations yield the three dimensional 
black holes \cite{BTZ,HWK}. 
The points $ r=r_0$ and $0$  become 
the event horizon and the singularity of the two dimensional black hole
whereas the outer and the inner horizon for the three dimensional black hole.

For the string model on 
$SL(2,R)$, there is a self-dual T-duality transformation (at semi-classical
level), ${\cal T}$, which
exchanges $ \theta_+ $ and $ \theta_- $ (i.e. $ \theta_R $ and $ -\theta_R$ ), 
or equivalently $ (r/r_0)^2 - 1/2 $ and $ 1/2 - (r/r_0)^2  $
\cite{NS}. 
This corresponds to the duality transformation to the two dimensional 
black hole geometry which exchanges the horizon and the singularity \cite{DVV}
and that to the three dimensional black hole geometry 
which exchanges the outer and inner horizon \cite{NS}.

Now let us consider the eigenvalue equation of the Laplace operator 
on $SL(2,R)$,
\eqabegin
   \lbb \frac{1}{\sqrt{-g}} \del_\mu \sqrt{-g} g^{\mu \nu} \del_\nu 
   -4j(j+1) r_0^{-2} \rbb \ \phi &=& 0 
  \comma
\eqaend
where $-j(j+1)$ is the Casimir. Physically, this is the Klein-Gordon equation
on $SL(2,R)$ and the Casimir is interpreted as mass or a coupling to the 
scalar curvature.(Notice that the scalar curvature is constant.)
Here, we rescale $ \theta_\pm $ as 
$ \tilde{\theta}_\pm = (\beta_\pm/2\pi r_0) \theta_\pm $ so that their 
imaginary periods associated with the Rindler space-like regions 
become $ \beta_\pm $, and make a separation of variables
$ \phi = \exp(i\omega_+ \tilde{\theta}_+ + i \omega_- \tilde{\theta}_-) 
\varphi(r) $.
The above equation then reduces to 
\eqabegin
  \lbb \Bigl\{ z (z-1) \del_z \Bigr\} ^2  + z(z-1) \Bigl\{ \frac{\nu_+^2}{z-1} 
   - \frac{\nu_-^2}{z} -j(j+1) \Bigr\} \rbb \varphi &=& 0
  \comma \label{Laplace}
\eqaend
with $ z = (r/r_0)^2 $ and 
\eqabegin
   \nu_\pm &=& - \frac{i}{4\pi} \beta_\pm \omega_\pm 
   \period
\eqaend
We find that the equation (\ref{Laplace}) is the same 
as the radial wave equation (\ref{appRN}) under the identifications
\eqabegin
  \nu_\pm & \leftrightarrow & \nu_\pm^{RN} \comma 
       \quad  \mbox{i.e.} \comma \quad 
  \beta_\pm \ \leftrightarrow \ \beta_{H,-}^{RN} \comma \nn \\
  j & \leftrightarrow &  j_l  \comma \qquad \qquad
    \abs{ \omega_\pm }  \ \leftrightarrow \abs { \omega }
  \period
\eqaend
This shows that the radial wave equation for the Reissner-Nordstrom 
black hole takes the $ SL(2,R) $ structure as we mentioned and that
the scalar propagation near the horizon is formally  the same as that 
in $SL(2,R)$.

The Laplacian on $SL(2,R)$ is expressed 
as $ J_{L,R}^i J_{L,R}^j \eta_{ij} $
using the left- or the right-$sl(2,R)$ currents .
So, we can rewrite (\ref{Laplace}) and hence (\ref{appRN}) 
in terms of the $sl(2,R)$ currents as
shown for the four and the five dimensional black holes 
\cite{CL,Hosomichi}.
In this expression, the $SL(2,R)$ structure becomes manifest.
This holds even for the Schwarzschild black hole.

Furthermore, the wave equations of tachyons or scalars in the two and 
the three dimensional 
black hole background also take the form  (\ref{Laplace}) 
in the entire geometries \cite{DVV,GLIS}.\footnote{
This is the case also for the three dimensional black string 
geometry \cite{3DBS}. 
}
This means that, for all the black holes discussed so far,
the propagation of scalars (or tachyons) near the horizon 
is governed by essentially the same
equation, namely, that for the $SL(2,R)$ geometry.\footnote{
The three dimensional black 
hole is not asymptotically flat. Because of this, the calculation 
of the absorption cross-section is different. 
However, by a certain definition of the asymptotic states at infinity, 
the CFT structure (\ref{CFT}) is obtained  \cite{BSS}. }
The solutions are then given by the hypergeometric functions in 
all the cases.
They are characterized by the geometric quantities 
associated with the Rindler-space regions, the horizons, the singularities
and/or infinity.
 
Next, we turn to the action of the T-duality transformation $ {\cal T} $ 
on the equation
(\ref{Laplace}). Note that $ (r/r_0)^2 -1/2 \ \to  \ 1/2 - (r/r_0)^2 $
corresponds to $ z \ \to 1-z$,  and $ \theta_R \ \to \ -\theta_R $ to 
$ \beta_R \equiv \nu_+ - \nu_- \ \to - \beta_R $ with 
$ \beta_L \equiv \nu_+ + \nu_- $ fixed.
Thus, from ${\cal T}^2 = 1$, (\ref{Laplace}) should be invariant under
\eqabegin
    z & \to & 1-z \comma \nn \\
   \beta_L & \to & \beta_L \comma \qquad \quad \beta_R \ \to \ - \beta_R
   \period 
\eqaend
We easily check that this is the case.  
Since the wave equations near the horizons are the same, 
this kind of symmetry is also 
common to all the cases we have discussed so far.
Moreover, in every case, this symmetry exchanges
the outer horizon and the inner horizon (or the singularity).
For the four and five dimensional black holes, 
this symmetry is nothing but the T-duality-like symmetry discussed in
\cite{CL}. Thus, we find that it corresponds to the actual T-duality
of the string model on $SL(2,R)$. 
%
%%%%%%%%%%%%%%%%%%%%%%%%%%%%%
\mysection{General argument}
%%%%%%%%%%%%%%%%%%%%%%%%%%%%%
%
%
In the previous section, we discussed
properties of the scalar propagation for the Reissner-Nordstrom and 
other black holes:
(i) The wave equations near the horizons have the $SL(2,R)$ structure. 
(ii) They also have a symmetry related to the T-duality  on $SL(2,R)$.
(iii) Their solutions are given by the hypergeometric functions and 
characterized by the information at the outer
horizons, the inner horizons (or the singularities) and infinity.
Typically, the periods of the imaginary time of the 
corresponding Rindler space appeared. (iv) In some parameter region, 
the absorption cross-sections at low energy take the form expected from a CFT. 

Then, a natural question is: how general are these properties ? In the
subsequent sections, we will show that they are common also to 
a large class of other spherically symmetric non-extremal
black holes and black $p$-branes. Furthermore, 
it turns out that most of the arguments does not depend on the details of
the geometries. Therefore, before moving on to explicit examples, it would 
be useful to discuss general features about the wave equations for 
 spherically symmetric black objects. Propagation of scalars in a general
class of spherically symmetric geometries has been discussed in \cite{Emparan}.

Here, we consider a spherically symmetric $p$-brane geometry 
in $D$-dimensions,
\eqabegin
  ds^2 &=& H(r)\lbb -f(r) dt^2 + dy^i dy^i \rbb + 
            R(r) \lbb h(r) dr^2 + r^2 d\Omega_{d+1}^2 \rbb 
  \comma
\eqaend
with $ i = 1,\cdots, p$ and $D=d+p+3$. We will deal with the asymptotically 
flat case in the following. 
This means that $f, h, H, R \ \to \ 1 $ as $ r \to \infty $. 
Since the geometries of black objects 
%(actually, all the geometries, to the author's knowledge) 
are usually expressed naturally by 
the harmonic functions $ 1/r^d $, we introduce a coordinate 
$x\equiv r^d$. In terms of $x$,  the radial 
equation of a scalar independent of $y^i$ becomes
\eqabegin
  \lbb \lb d K \del_x \rb^2  + U(x;\omega) \rbb \varphi &=& 0
   \comma \label{genericpB}
\eqaend
where
\eqabegin
  K(x) & = &  x^2 H^{\frac{p}{2}} R^{\frac{d+1}{2}} (-g_{tt}/g_{rr})^{\half}
     \comma \quad 
  U(x;\omega) \ = \ x^2 H^p R^{d} \lb R x^{\frac{2}{d}} \omega^2
                    - f H \Lambda \rb
  \period
\eqaend 

First, we note that the kinetic term vanishes where $K=0$ and 
the differential equation becomes singular there. 
Since $ (dr/dt)^2 = (-g_{tt}/g_{rr}) $ 
for the light-like geodesics, this typically occurs 
at the horizons and singularities of the geometry.

Next, as is the case for the geometries discussed in section 2, 
suppose that $K(x)$ has the form $ \gamma (x-x_+)(x-x_-)$ with 
$ \gamma $ a non-vanishing constant. 
Then, we find that $x=x_\pm$ become
regular singular points if $ U(x;\omega)$ is analytic there 
with respect to $x$. Introducing a new coordinate $ y = 1/x $, the equation
(\ref{genericpB}) becomes
\eqabegin
  \lbb \Bigl\{ \gamma d (1-x_+ y)(1-x_- y) \del_y \Bigr\}^2  
   + \frac{1}{y^2}  H^p R^d \lb R y^{-2/d} \omega^2 
     - f H \Lambda \rb  \rbb 
  \varphi &=& 0 
  \period
\eqaend
Taking the asymptotic flatness into account, we see that $y=0 $ 
$(x=\infty)$ is 
also a singular point but it is irregular because of the term $ y^{-2/d}$.

From the above observation, we find that, by any approximation 
which makes the potential term $U(x;\omega)$ an at most quadratic 
polynomial of $x$, the wave equation (\ref{genericpB}) becomes a differential 
equation which has only three regular singular points.
The approximated solutions are then expressed by the hypergeometric 
functions. We can always make this kind of approximation 
at least in the neighborhood of a point by expanding $U(x;\omega)$ if 
$U(x;\omega)$ is regular there.  
Almost all the approximations in the literature which lead to 
the absorption cross-sections of the form (\ref{CFT}) fall into 
this category.
Note that $ d = 1 $ and $ 2 $ cases ($D = 4$ and $5$ for black hole geometries)
are special in that $ r^2 = x^{2/d} $ in $U(x;\omega)$ is a monomial of $x$.
Because of this fact, one can use approximations of the type used, 
e.g., in \cite{MS,KM}, which are impossible for $d>3$.  

The hypergeometric functions are characteristic of $SL(2,R)$
symmetry; they are the matrix elements of $SL(2,R)$ representations. 
So, we expect that the $SL(2,R)$ structure discussed 
in section 2 also appear even in the above general situation. 
In fact, we can confirm this as follows. Let us denote the approximated
potential by
\eqabegin
  \tilde{U}(x;\omega) &=& (x-x_+)(x-x_-) \lbb \frac{\tilde{U}_\infty}{\mu^2} 
         + \frac{\tilde{U}(x_+)}{\mu(x-x_+)} 
        - \frac{\tilde{U}(x_-)}{\mu(x-x_-)} \rbb
   \comma
\eqaend
with $ \mu \equiv x_+ - x_- $. Then, in terms of $ z = (x-x_-)/\mu$, 
the wave equation becomes
\eqabegin
  \lbb \Bigl\{ z(z-1) \del_z \Bigr\} ^2 +  z(z-1) 
    \Bigl\{ \frac{\bar{\nu}_+^2}{z-1}
   - \frac{\bar{\nu}_-^2}{z} + 
  \frac{\tilde{U}_\infty}{(d\gamma\mu)^2} \Bigr\} \rbb
   \varphi &=& 0
   \comma
\eqaend 
where $ \bar{\nu}_\pm^2 = - \tilde{U}^2(x_\pm)/(d\gamma\mu)^2 $.
This is the same as the equation (\ref{Laplace}) under appropriate 
identifications of the parameters. Therefore, we find that the above equation 
indeed possesses the $SL(2,R)$ structure and the T-duality-like symmetry 
discussed in section 2.4. We remark that a differential equation with 
three regular singular points does not always take the form 
(\ref{Laplace}); for this to be the case, the exponents associated 
with two singular points, 
$\nu_{1(2)}$ and $\nu_{1(2)}'$, need to satisfy $ \nu_{1(2)} = -\nu_{1(2)}'$.

Under appropriate conditions on the parameters,
the original wave equation (\ref{genericpB}) can be simplified 
to the Bessel-type
equation like (\ref{Bessel}) near infinity.
So, the calculation of the absorption cross-section can be performed 
similarly to section 2.2.
In this picture, the matching procedure
is formally the same as the matching of the wave function on $ SL(2,R)$
and that in flat Minkowski space-time. 
We would then find the CFT structure of the cross-section in the parameter
region where the non-trivial $\omega$-dependence is obtained.

In this section, we saw that (i) the wave equation for a black object
may have singular points typically at the horizons, singularities and infinity,
(ii) if the function $K(x)$ in the kinetic term takes the form
$ \gamma (x-x_+)(x-x_-) $, the wave equation can be approximated 
by the equation of the form (\ref{Laplace}), 
(iii) then, the properties discussed in section 2 may hold even 
in this general situation.

In the following sections, we will confirm that these are actually realized
for a large class of black objects. Moreover, we will find that 
the potential terms proportional to $ \Lambda $ always become 
of the form $(x-x_+)(x-x_-) \Lambda$.
Consequently the only difference is encoded in the potential
term corresponding to $V_{RN}$.
%
%%%%%%%%%%%%%%%%%%%%%%%%%%%%%%%%%%%%%%
\mysection{Dyonic dilaton black holes}
%%%%%%%%%%%%%%%%%%%%%%%%%%%%%%%%%%%%%%
%
The Reissner-Nordstrom black holes belong to a class of 
black holes with two
regular horizons. In this section, we deal with another class of 
such black holes, namely, the dyonic dilaton black holes by Gibbons and 
Maeda \cite{GM}. In almost the same way as in section 2, 
we will find that the properties of the scalar propagation discussed so far 
hold for these black holes. 
\mysubsection{Space-time geometry}
The metric of the dyonic dilaton black holes in $D$-dimensions 
is given by\footnote{
Our notation is somewhat different from that in \cite{GM}. The relation is:
$ \eta \leftrightarrow r^d $, $ \eta_{1,3} \leftrightarrow Q_{1,2} $. }
\eqabegin
  ds^2_{GM} &=& - f_{GM}(r) \lambda_{GM}^d (r) dt^2 + \lambda_{GM}^{-1}(r)
   \lbb f_{GM}^{-1}(r) dr^2 + r^2 d\Omega_{d+1}^2 \rbb
   \comma 
\eqaend
where $ d = D-3 $;
\eqabegin
    f_{GM}(r) &=& \lb 1 - \frac{\eta_0}{r^d} \rb 
                  \lb 1 + \frac{\eta_0}{r^d} \rb 
   \comma  \quad 
     \lambda_{GM}^{-d} \ = \   \lb 1+\frac{Q_1}{r^d}\rb^{\alpha_1} 
                \lb 1+\frac{Q_2}{r^d}\rb^{\alpha_2}  
  \comma \nn \\
   \alpha_1 &=& \frac{2g_{2}}{g_2+g_{D-2}} \comma \qquad 
   \alpha_2 \ = \ \frac{2g_{D-2}}{g_2+g_{D-2}} \comma \qquad 
    g_2 g_{D-2} \ = \ d
  \ ; 
\eqaend
$ g_{2(D-2)} $ is a parameter related to the coupling between 
the dilaton and the two (($D-2$)-) form. Without loss of generality, 
we can set 
$\alpha_1 \geq \alpha_2 $. Also, we concentrate on the case 
 $Q_{1,2} > \eta_0 > 0 $ in which the geometry has two regular horizons
at the points  $r^d = \pm \eta_0 $ and the singularity 
at $ r^d = - \min \{Q_1, Q_2 \}$. When $ Q_1 =  Q_2$, the geometry reduces 
to that of the Reissner-Nordstrom black hole. 
Similarly to section 2, we introduce 
$ z \equiv (r^d + \eta_0)/\mu $; $\mu =  2 \eta_0 \equiv r_0^2$. 
The extremal limit
corresponds to $ \mu \to 0 $. In terms of $z$, the metric becomes
of the form (\ref{RNV})
with 
\eqabegin
  V_{GM} (z) &=& \lb \frac{r}{r_0}\rb ^{2(d+1)} 
    \lambda^{-(d+1)}_{GM} 
   \ = \ \lmb \lb  z-\half  + \hat{Q}_1 \rb^{\alpha_1} 
   \lb  z-\half 
    + \hat{Q}_2 \rb ^{\alpha_2} \rmb^{1+\frac{1}{d}}
  \comma 
\eqaend
instead of $ V_{RN}$, and $ \hat{Q}_{1,2} = Q_{1,2}/\mu $. 
Thus we readily obtain the 
periods of the imaginary time associated with the near horizon 
geometries,
\eqabegin
  \beta_H^{GM} &=&  \frac{4\pi}{d} r_0 V_{GM}^{\half}(1)
   \ = \ \frac{4\pi}{d} r_0 
        \lmb \lb \hat{Q}_1 +\half \rb^{\alpha_1}
             \lb \hat{Q}_2+\half \rb^{\alpha_2}\rmb^{\frac{d+1}{2d}}
   \comma \nn \\
   \beta_-^{GM} &=&  \frac{4\pi}{d} r_0 V_{GM}^{\half}(0)
   \ = \ \frac{4\pi}{d} r_0 
          \lmb \lb \hat{Q}_1 -\half \rb^{\alpha_1}
               \lb \hat{Q}_2-\half \rb^{\alpha_2}\rmb^{\frac{d+1}{2d}}
  \period
\eqaend
\mysubsection{Wave equation and absorption cross-section}
Now we turn to the discussion on the propagation of a minimally 
coupled massless scalar. It is easy to find that 
the wave equation is obtained from (\ref{weqRN}) simply 
by the replacement $ V_{RN} \to V_{GM} $. We see that the $V_{GM}$ 
(or $\lambda_{GM}$)-dependence disappears in the kinetic term 
and the potential term proportional to $\Lambda$. Thus the wave 
equation reduces to the form discussed in section 3.

We then examine the wave equation near the horizon.
In this region, we can approximate $ V_{GM} $ by its value 
at the outer horizon;\footnote{
It turns out that, even if we include the linear term in the expansion, it is 
irrelevant to the final result of the cross-section as in section 2.}
\eqabegin
  V_{GM} (z) &\sim & \tilde{V}_{GM} (z) \ \equiv \ V_{GM} (1)
  \period
\eqaend 
Since
\eqabegin 
  \del_z V_{GM} & = & -\lb 1 +\frac{1}{d} \rb  V_{GM}(z) 
      \lbb \frac{\alpha_1}{(z-1/2) +\hat{Q}_1} 
   + \frac{\alpha_2}{(z-1/2) +\hat{Q}_2} \rbb
 \comma
\eqaend
$ \tilde{V}_{GM} $ is a good approximation for 
$ \abs{ z-1 } \ll \min \{ \hat{Q}_{1}, \hat{Q}_{2} \} $.
By this approximation, the total potential term  becomes
a quadratic polynomial of $z$ and the wave equation again
takes the form (\ref{Laplace}). As a result, we find that 
it possesses the $SL(2,R)$ structure and the T-duality-like symmetry. 
In this case, the exponents associated with the singular points are
\eqabegin
  \nu^{GM}_\pm & = & - \nu^{GM'}_{\pm} \ =  
    - i \frac{r_0 \omega}{d} \tilde{V}^{\half}_{GM} 
     \ = \ -  \frac{i}{4 \pi} \beta_H^{GM} \omega \comma \nn \\
   \nu^{GM}_\infty & = & 1 - \nu^{GM'}_{\infty} \ = \ - j_l
  \period 
\eqaend

The wave equation in the far region $r^d \gg Q_{1,2}$ simplifies 
to (\ref{Bessel}). 
Therefore, the calculation of the absorption cross-section is identical with 
that in section 2.2, and hence the cross-section is given by 
Eq.(\ref{abscross}) with $ \beta_H^{GM} $ instead of $\beta_H^{RN}.$

As in the previous section, we still need to check the range of validity
of the result. Since the procedure is similar to that in section 2.3, 
we will just give the points. First, we focus on $\Lambda \neq 0$.
For the approximation to be valid at the matching point $r=r_m$, 
we require that $ r_m^d \gg Q_{1,2}, \eta_0 $ and $ r_m \omega \ll 1 $ 
should hold.
Under these conditions, $ \abs{ \nu_\pm^{GM} } \sim \beta_H^{GM} \omega $ 
can be of order unity if $ \hat{Q}_{1,2} $ are large enough, 
and then the non-trivial $\omega$-dependence appears. 
Although, in some parameter region, 
$ \abs{ \nu_\pm^{GM} } \sim {\cal O}(1)$ even when one of $\hat{Q}_{1,2}$ 
is small, 
let us concentrate on the case $ \hat{Q}_1 \geq \hat{Q}_2 \gg 1$ and  
$ \abs{ \nu_\pm^{GM} } \sim {\cal O}(1)$ for simplicity.  
For $\abs{ z-1 } \simg \hat{Q}_2 $, $\tilde{V}_{GM}$ is not 
a good approximation to $ V_{GM} $. So, we need to require that 
$z(z-1)\Lambda$ term dominates 
$(r_0 \omega)^2 V_{GM}$ and $(r_0 \omega)^2\tilde{V}_{GM}$ there
as in (\ref{nearcond}).
After some algebra, we find that this is satisfied without any further
conditions. Therefore, we confirm that the absorption cross-section 
has the CFT structure for 
\eqabegin
  && Q_i^{1/d} \omega \ll 1 \comma \qquad \hat{Q}_i \gg 1 \comma 
    \qquad \abs{ \nu_{\pm}^{GM} } \sim {\cal O} (1)
   \period
\eqaend
Finally, $\Lambda = 0$ case gives (\ref{arealaw}) as before. 
%
%%%%%%%%%%%%%%%%%%%%%%%%%%%%%%%%%%%%%%%%%%%%%%%%%
\mysection{Charged black holes in string theory}
%%%%%%%%%%%%%%%%%%%%%%%%%%%%%%%%%%%%%%%%%%%%%%%%%
%
The black holes in section 2 and 4 had two regular horizons.
In this section, we discuss charged black holes in string theory which 
have only one horizon in a generic case. We will find that the propagation 
of scalars is similar to the previous cases; the wave equations near the 
horizons have the properties associated with $SL(2,R)$ and 
the absorption cross-sections take the CFT structure 
under certain conditions of validity.
\mysubsection{Space-time geometry}
A class of $D$-dimensional charged black holes in string theory has 
the Einstein metric \cite{Maldacena},\cite{Youm}-\cite{dAS},
\eqabegin
  ds_{CB}^2 &=& -  f_{CB}(r) \lambda^{d}_{CB}(r) dt^2 + \lambda_{CB}^{-1} (r) 
                  \lbb  f_{CB}^{-1} (r) dr^2 + r^2 d\Omega_{d+1}^2 \rbb
  \comma \label{CB}
\eqaend
where $d=D-3$ and 
\eqabegin
  \lambda_{CB}^{-(d+1)} (r) &=& \prod_{i=1}^{n} \lb 1+ \frac{Q_i}{r^d} \rb
   \comma \qquad f_{CB}(r) \ = \ 1 -\frac{\mu}{r^d}
  \period
\eqaend 
For later use, we define a parameter
\eqabegin 
  \xi_{CB}  & = & 1 + \frac{1}{d} - \frac{n}{2} 
  \period
\eqaend
$ \xi_{CB}  = 0 $ cases, i.e., $D=5$, $n=3$ and $D=4$, $n=4$, 
correspond to the five and four 
dimensional black holes extensively studied in relation to  $D$-branes.
$ n = 2 $ cases include the black holes closely related to fundamental
strings \cite{Maldacena}. 
The scalar propagation for $\xi_{CB} \neq 0$ has been 
discussed in \cite{dAS,Emparan2}.
The geometry has a horizon at $ r^d = \mu \equiv r^d_0 $ and, 
in a generic case $(\xi_{CB}  \neq 0)$, 
a singularity develops at $r=0$. $\mu \to 0$ is the extremal limit.
Introducing $ z = r^d/\mu$, the metric takes the form (\ref{RNV})
with 
\eqabegin
   V_{CB}(z) &=& \lb \frac{r}{r_0} \rb ^{2(d+1)} \lambda_{CB}^{-(d+1)} \ = \ 
   z^{2 \xi_{CB} } \prod_{i=1}^n \lb z + \hat{Q}_i \rb
\eqaend
instead of $ V_{RN}$, and $\hat{Q}_i =  Q_i/\mu$.
The period of the imaginary time associated with the near-horizon geometry
is \cite{CT2}
\eqabegin
  \beta_H^{CB} &=& \frac{4\pi}{d} r_0 V_{CB}^{\half}(1)
   \ = \ \frac{4\pi}{d} r_0 \prod_{i=1}^n (1+\hat{Q}_i)^{\half}
  \period
\eqaend
When $\xi_{CB}  = 0$, $r=0$ becomes a regular horizon and 
the corresponding period $\beta_-^{CB} $ is obtained by the replacement 
$ V_{CB} (1) \to V_{CB} (0)$.  For $\xi_{CB}  \neq 0$, we do not have 
$ \beta_-^{CB} $
because $ r = 0 $ is a singular point. We can see this also from 
the fact that $ V_{CB} (0) $ is not a non-vanishing constant.
\mysubsection{Wave equation and absorption cross-section}
Now we consider the propagation of a minimally coupled massless
scalar. We readily find that the radial wave 
equation takes the form (\ref{weqRN}) with $ V_{CB} $ instead of $ V_{RN}$.
The $V_{CB}$ (or $\lambda_{CB}$)-dependence 
again disappears in the kinetic term and the potential term proportional to 
$\Lambda$.  The kinetic term vanishes at $ z=1$ and $z=0$. $z=1$ 
corresponds to the horizon as in the previous cases, but in a generic case
$z=0$ corresponds to the singularity instead of the inner horizon.

First, we study the wave equation near the horizon.
To make the discussion definite, 
we will concentrate on the case $ \xi_{CB} > 0 $ in what follows.
Approximating $V_{CB}$ by the expansion around the horizon, we find that
the approximated equation possesses the $SL(2,R)$ structure and the 
T-duality-like symmetry. However, it turns out that this approximation does
not give the non-trivial frequency dependence of the absorption cross-section;
under the condition of validity, the 
cross-section reduces to the leading term in (\ref{abscrossLR}).
This is due to the singular nature at $ r=0$. In fact, since 
\eqabegin
  \del_z V_{CB} &=& V_{CB} \lb \frac{2\xi_{CB}  }{z} + \sum_i 
               \frac{1}{z+\hat{Q}_i} \rb
  \comma 
\eqaend
the expansion gives a good approximation only for $ \abs{ z-1 } \ll 1 $
when $\xi_{CB} \neq 0$.
Thus, in the region $ z \sim 0 $ $ (r \sim 0)$ where the other 
potential term $z(z-1) \Lambda$ vanishes, 
the error becomes significant so that  the validity of the approximation 
is not assured in the parameter region we are interested in.

Therefore, we will take a different approximation here;
\eqabegin
   V_{CB}(z) &\sim & \tilde{V}_{CB}(z) \ = \ z \prod_i (1+\hat{Q}_i)
   \period
\eqaend 
This gives a good approximation where $z(z-1) \Lambda$ becomes vanishing.
Moreover, for $ \xi_{CB} = 1/2 $, the approximation is valid
even for $ \abs{ z } \ll \min \{ \hat{Q}_i \} $ because the error 
$ V_{CB} - \tilde{V}_{CB} $ is of order $ z/\hat{Q}_i $.  

The wave equation near the horizon then takes the
form (\ref{Laplace}) with the exponents
\eqabegin
  \nu^{CB}_+ & = & - \nu^{CB'}_{+} \ = \ 
    - i \frac{r_0 \omega}{d} \tilde{V}^{\half}_{CB}(1) 
     \ = \ -  \frac{i}{4 \pi} \beta_H^{CB} \omega \comma \nn \\
  \nu^{CB}_- & = & - \nu^{CB'}_{-} \ =  \
   - i \frac{r_0 \omega}{d} \tilde{V}^{\half}_{CB}(0) \ = \ 0
     \comma \\ 
   \nu^{CB}_\infty & = & 1 - \nu^{CB'}_{\infty} \ = \ - j_l \nn
  \period 
\eqaend
Thus we confirm that the wave equation has the properties 
related to $SL(2,R)$.

The calculation of the absorption cross-section is the same as in 
the previous cases. 
The wave equation in the far region $ r^d \gg Q_{i}, \mu $ reduces to the 
Bessel-type equation and the cross-section is expressed by 
(\ref{abscrossLR}) with
\eqabegin
  && \beta_L^{CB} = \beta_{R}^{CB} = \beta_H^{CB}
  \comma 
\eqaend
instead of $ \beta_{L,R,H}^{RN}$.

We then turn to the discussion on the range of validity. Without loss of 
generality we set $ \hat{Q}_1 = \max \{ \hat{Q}_i \}$ and, for simplicity, 
we deal with $\hat{Q}_i \gg 1$. We also set the matching point $ r=r_m $
$ (z=z_m)$  large enough so that $V_{CB}  \simg   \tilde{V}_{CB}$ 
there.\footnote{
For $\xi_{CB} < 1/2$, we can choose the matching point so that 
$\tilde{V}_{CB} \gg V_{CB} $.}
This yields the condition
\eqabegin 
      z_m^{1+2/d} & \simg & \prod \hat{Q}_i 
     \period \label{VtilV}
\eqaend
First, we discuss $\Lambda \neq 0$ case.
For the approximation to be valid at the matching point,  
 we require that $ r_m^d \gg Q_i, \mu $; $ r_m \omega \ll 1 $ 
and (\ref{nearcond})
with $ V_{CB} $ instead of $ V_{RN} $ should hold. This gives 
\eqabegin
  z_m \gg \hat{Q}_i \gg  1 \comma && 
   r_m \omega = r_0 \omega z_m^{1/d} \ll 1
  \period \label{cond}
\eqaend
Let $ r_0 \omega = \epsilon z_m^{-1/d} $ and $ z_m = L \hat{Q}_1 $ with
$ L, \epsilon^{-1} \gg 1 $
so as to satisfy these conditions. Then we have
\eqabegin
&& \abs{ \nu_{+}^{CB} } \sim \beta_H^{CB} \omega 
  \sim r_0 \omega  \prod \hat{Q}_i^{1/2} \sim \epsilon L^{-1/d } 
    \hat{Q}_1^{-1/d} \prod \hat{Q}_i^{1/2} 
 \comma
\eqaend 
This indicates that, for $ \xi_{CB} < 1 $,  
$ \abs{ \nu_+^{CB} } $ 
can be of order unity if the charges are sufficiently large, 
but it is impossible for  $ \xi_{CB}  \geq 1  $ 
in this approximation.

Next, we examine the error from $ z < z_m $. We consider the 
case $ \abs{ \nu_+^{CB} } \sim {\cal O}(1) $. For 
$ \xi_{CB} \neq 1/2 $, $ \tilde{V}_{CB} $
well approximates $V_{CB}$ near $ z= 0,1$.
Thus the requirement for the validity is that 
the condition (\ref{nearcond}) with $V_{CB}$
holds except near these points, namely, at the scales $z \sim \hat{Q}_i$, 
$z \sim {\cal O} (1)$ and
$ z \sim -1/2 $. This is satisfied if
\eqabegin 
     \hat{Q}_i \gg 1 \comma && 
   d^2 \abs{ \nu_{+}^{CB} } ^2 \hat{Q}_i^{-2(1-\xi_{CB})} 
     \prod_k^{n} \lb 1+ \hat{Q}_i/\hat{Q}_k \rb \ll \Lambda 
    \qquad ( i = 1 , \cdots , n )
   \comma \label{condQ} \\
    && \qquad \qquad \Lambda \gg  d^2 \abs{ \nu_{+}^{CB} } ^2 \period
    \label{condLambda}
\eqaend  
For $ \xi_{CB} = 1/2 $, 
the approximation is better and the requirement 
is that (\ref{nearcond}) with $V_{CB}$ holds for 
$ z \sim \hat{Q}_i $. This gives the condition (\ref{condQ}).
Therefore, the non-trivial frequency dependence in 
(\ref{abscrossLR}) with $\beta^{CB}_{L,R,H}$ is obtained 
in the parameter regions 
(i) (\ref{VtilV}), (\ref{cond}), (\ref{condQ}) and 
$ \abs{ \nu_+^{CB} } \sim {\cal O} (1) $ for $ \xi_{CB} = 1/2 $, 
(ii) (i) plus (\ref{condLambda}) for 
$ 0< \xi_{CB} < 1 $, $ \xi_{CB} \neq 1/2 $.
In these parameter regions,  the absorption cross-section has the 
CFT structure.

For $\Lambda = 0$, we obtain (\ref{arealaw}).

As discussed in section 3, $r^2 \propto z^{2/d}$ becomes
a positive integral power of $z$ for $D=4$ and $5$ 
and one can utilize more precise 
approximations of the type used in, e.g., \cite{MS,KM}. 
In fact, the results for $ \xi_{CB} = 1/2 $ ($D=4, n=3$ and $D=5, n=2$)
in this section can be compared with \cite{KM} and 
we find agreement; our results of the absorption cross-sections
reduce to those in \cite{KM} in the corresponding parameter region 
with the smallest charge vanishing.
This agreement holds even for $ \Lambda = 0 $.
%\footnote{
%In the `dilute gas region' \cite{MS,HSM} of the geometry (\ref{CB}),
%it has been shown that, when the smallest charge is non-zero, 
%the $s$-wave absorption cross-sections
%take the factorized form as in (\ref{abscrossLR}) only if $ \xi_{CB} = 0$. 
%There is no contradiction between this result and the factorized from 
%of the absorption cross-sections
%in \cite{KM} with the vanishing smallest charge.}
%
This suggests
that more elaborated approximations may also give (\ref{abscrossLR})
with $\beta^{CB}_{L,R,H}$ and the range of its validity 
may be wider than discussed in this section. 
%
%%%%%%%%%%%%%%%%%%%%%%%%%%%%%%
\mysection{Black $p$-branes}
%%%%%%%%%%%%%%%%%%%%%%%%%%%%%%
%
We have seen that the propagation of scalars 
possesses interesting features for various 
black hole geometries. 
As a final example, we discuss a class of non-extremal black $p$-branes.
We will find that the properties discussed so far hold also in 
this case.
\mysubsection{Space-time geometry}
It turns out that 
we can deal with all the $p$-brane solutions in \cite{DLP,Guven}
in the same way. However, to make the discussion definite and concise, 
we will concentrate on the class of the $D$-dimensional solutions 
with the Einstein metric 
\cite{DLP},\footnote{
We basically follow the notation in \cite{KT}. }
\eqabegin
 ds_{pB}^2 &=& H^\alpha (r) \lb H^{-N}(r) 
    \lbb -f_{pB}(r) dt^2 + dy^i dy^i \rbb
        + f^{-1}_{pB} (r) dr^2 + r^2 d\Omega_{d+1}^2 \rb
   \comma \nn \\
  && H(r) \ = \ 1 + \frac{Q}{r^d} \comma 
     \qquad f_{pB}(r) \ = \ 1 - \frac{\mu}{r^d} \comma \\
  && \alpha \ = \ \frac{p+1}{d+1} N \comma 
    \qquad N \ = \ 4\lbb g_{D-2}^2 + 2d\frac{p+1}{d+1} \rbb^{-1}
   \comma \nn
\eqaend
where $ i = 1, \cdots, p$ and $D=d+p+3$. $g_{D-2}$ is a parameter related to 
the coupling between the dilaton and the $(D-2)$-form. 
This includes the $p$-brane solutions in \cite{pbranes}.
$g_{D-2}=0$ corresponds
to the constant dilaton. In this (non-dilatonic) case, 
the near extremal entropy scales as that of the massless ideal gas in 
$(p+1)$-dimensional world volume \cite{KT}. The geometry has a horizon 
at $ r^d = \mu \equiv r_0^d$. When $\mu = 0$, the geometry becomes extremal.
In a generic case, $r=0$ is the location of the singularity. When $N$ is an
integer, the extremal solution becomes supersymmetric. 

Introducing $ z = r^d/\mu $, the metric becomes
\eqabegin
  ds_{pB}^2 &=& H^{ \frac{dp\alpha}{(d+1)(p+1)} } \lmb 
     \ V_{pB}^{-\frac{d}{d+1} } 
         \lbb -z(z-1) dt^2 + z^2 dy^i dy^i \rbb  \right. \label{pBV} \\
    && \left. \qquad \qquad  \qquad \qquad \qquad 
         + \ r_0^2 V_{pB}^{ \frac{1}{d+1} } 
   \biggl[  \frac{1}{d^2} \frac{dz^2}{z(z-1)} + d\Omega_{d+1}^2 \biggr] 
   \rmb
  \comma \nn 
\eqaend
with
\eqabegin
  V_{pB} (z) &=& \lb \frac{r}{r_0} \rb ^{2(d+1)} H^{N} \ = \ 
      z^{2\xi_{pB}} \lb z+\hat{Q} \rb^{N}
   \period
\eqaend
Here we have defined $ \hat{Q} = Q/\mu $ and 
\eqabegin
  \xi_{pB}  &=& 1 + \frac{1}{d} - \frac{N}{2} \ \geq \ 0
    \period
\eqaend 
Except for the overall factor and the $p$-brane part, $ z^2 dy^i dy^i $, 
the above metric is the same as (\ref{RNV}) with $V_{pB}$ instead of $V_{RN}$.
Then the period of the imaginary time associated
 with the horizon is \cite{DLP,KT} 
\eqabegin
  \beta_H^{pB} &=& \frac{4\pi}{d} r_0 V_{pB}^{\half}(1)
   \ = \  \frac{4\pi}{d} r_0 (1+\hat{Q})^{\frac{N}{2}}
  \period
\eqaend
If $r=0$ is an inner horizon, we get the corresponding period similarly.
\mysubsection{Wave equation and absorption cross-section}
The $p$-brane geometry (\ref{pBV}) is very similar to 
the previous black hole cases, and
 so is the discussion on the propagation of scalars. 
After some calculation, we find that, in the radial wave equation, 
the $H$-dependence cancels 
in the kinetic term and the term proportional to $\Lambda$.
The wave equation is then given by (\ref{weqRN}) 
with $V_{pB}$.\footnote{The kinetic term 
of the wave equation takes the same form also for the $p$-brane solutions 
in \cite{LPX}
}
Comparing $ V_{pB}$ with $V_{CB}$, we find that, when $N$ is an integer, 
the wave equation for $ V_{pB} $ is the same as the equation for $ V_{CB} $ 
with $ \hat{Q}_i = \hat{Q} $ and $ n = N $.

Let us move on to the detail.
First, we discuss the special case $ \xi_{pB} = 0 $.
In this case, the entropy is non-vanishing in the extremal limit 
\cite{DLP,KT}, and the geometry for $ p=0 $ becomes 
that of the Reissner-Nordstrom black hole.
We see that, only when $ D=4$ and $5$, 
$ N $ is an integer and the extremal Reissner-Nordstrom black 
holes become supersymmetric. 

The discussion for an integral $N$ reduces to that
for $\xi_{CB} = 0$ case with the equal charges in the previous section. 
In this case, we can use the analysis in the literature or in section 2.
Furthermore, because
\eqabegin
   \del_z V_{pB} &=& V_{pB} \lb \frac{2\xi_{pB} }{z} + \frac{N}{z+\hat{Q}} \rb
   \comma
\eqaend
the expansion of $V_{pB}$ gives a good approximation 
in the region $ \abs{ z } \ll \hat{Q} $ even for a non-integral $N$.
This is analogous to 
the black holes with regular inner horizons in section 2 and 4, 
and similar analysis is possible. Consequently, the near-horizon 
wave equation has the properties related to $ SL(2,R)$ and 
the absorption cross-section has the CFT structure in every case
with $ \xi_{pB} = 0 $.

Next, we consider $ \xi_{pB}  \neq 0 $ case.
For the same reason as in section 5, we approximate the wave equation by
\eqabegin
   V_{pB} (z) \ \sim  \ \tilde{V}_{pB}(z) &=& z (1+\hat{Q})^{N}
    \period
\eqaend
Although the possible values of 
$ \xi_{CB} $ and $\xi_{pB}$ are different, 
it then turns out that the discussion on the scalar propagation 
is almost the same.
We can obtain the results for the $p$-branes from those in the 
previous section simply by 
the replacement $ \xi_{CB} \to \xi_{pB} $, $\hat{Q}_i \to \hat{Q}$,
$ n \to N $ and so on.

Explicitly, the approximated wave equation near the horizon is given by 
(\ref{Laplace}) with the exponents
\eqabegin
  \nu^{pB}_+ & = & - \nu^{pB'}_{+} \ = \ 
    - i \frac{r_0 \omega}{d} \tilde{V}^{\half}_{pB}(1) 
     \ = \ -  \frac{i}{4 \pi} \beta_H^{pB} \omega \comma \nn \\
   \nu^{pB(')}_- & = & 0 \comma \qquad  
   \nu^{pB}_\infty \ = \ 1 - \nu^{pB'}_{\infty} \ = \ - j_l 
    \period
\eqaend
We can confirm the properties associated with  $SL(2,R)$
regarding this wave equation.

As for the absorption cross-section, 
it is obtained from (\ref{abscrossLR}) with 
\eqabegin
  && \beta_L^{pB} = \beta_R^{pB} = \beta_{H}^{pB}
   \comma  
\eqaend
instead of $\beta_{L,R,H}^{RN}$.
The range of validity is examined in the same way.
We choose the matching point so that 
$ V_{pB} \simg \tilde{V}_{pB} $ there. This yields the condition
$ z_m^{1+2/d} \geq \hat{Q}^N $. For $ \Lambda \neq 0$, 
the matching procedure is valid if
$ z_m \gg \hat{Q} $ and $ r_m \omega = r_0 \omega z_m^{1/d} \ll 1 $.
Thus $ \abs{ \nu_+^{pB} } $ can be of order unity 
for sufficiently large $ \hat{Q} $ and $ \xi_{pB}  < 1 $.
The estimation of the error from $ z < z_m $ leads to the conditions
(i) $ d^2 \abs{ \nu_{+}^{pB} } ^2 \gg \Lambda $; $ \hat{Q} \gg 1 $  
for $ \xi_{pB} \neq 1/2 $ and (ii) $ \hat{Q} \gg 1 $ for 
$ \xi_{pB} = 1/2 $. 
Therefore, the non-trivial $\omega$-dependence is obtained 
in the parameter regions  
\eqabegin
  && \xi_{pB} < 1 \comma \ \xi_{pB} \neq \half \ ; \qquad    
    Q^{1/d} \omega \ll 1 \comma \qquad 
     \hat{Q} \comma \Lambda \gg 1 \comma 
    \qquad \abs{ \nu_+^{pB} } \sim {\cal O}(1) \comma \nn \\
  && \xi_{pB} = \half \ ; \qquad \qquad   \qquad \ 
    Q^{1/d} \omega \ll 1 \comma \qquad \hat{Q} \gg 1 \comma
     \qquad \quad  \ \! \abs{ \nu_+^{pB} } \sim {\cal O}(1)
   \period
\eqaend 
The absorption cross-section has the CFT structure there.
%
%
%%%%%%%%%%%%%%%%%%%%%%
\mysection{Summary and discussion}
%%%%%%%%%%%%%%%%%%%%%%
%
In this paper, we discussed the propagation of minimally coupled massless
scalars in various non-extremal black hole and $p$-brane geometries. 
We showed that 
some of the properties known about a certain class of
 non-extremal four and five dimensional 
black holes hold very generally: (i)
The radial wave equations near the horizons approximately take the same form 
as the eigenvalue 
equation of the Laplace operator on $SL(2,R)$. (ii) The solutions there 
are characterized by the information at the outer horizons, the inner horizons
(or the singularities) and infinity and they are expressed by 
the hypergeometric functions. Typically, 
the periods of the imaginary time 
associated with the near-horizon geometries appear in those solutions.
(iii) The wave equations 
have a symmetry related to the T-duality of the 
string model on $ SL(2,R) $.
(iv) The absorption cross-sections at very low energy 
take the form expected form a CFT in some parameter region. 
We saw that the above features of the near-horizon wave
equations are valid irrespectively of extremality and supersymmetry. 
The properties (i)-(iii) were  
also common to the two dimensional $SL(2,R)/U(1)$ black holes and the 
three dimensional BTZ black holes.

For the four and five dimensional black holes, the above properties
hold  also 
in the rotating cases \cite{MS2,CL,early} and for particles with higher spins
\cite{Gubser,Hspins,Hosomichi}.
Taking this into account, we expect that it is possible to extend 
the argument in this paper to those cases. Moreover, 
the structures of the geometries were very similar in terms of the quantities
corresponding to $ V_{RN} $ and the metrics could be written as (\ref{RNV})
in all the cases. This suggests a possibility
that, under quite general assumptions, black objects should possess 
the properties such as (i)-(iv). Also, by more elaborated approximations, 
it may be possible to 
extend the parameter regions in which the non-trivial frequency dependence
of the cross-sections is obtained.
 For example, $ V_{CB [pB ] } $ is well
approximated near the horizon by $ z^2 \prod (1+\hat{Q}_i) $ 
$ [ z^2 (1+\hat{Q})^N ] $ for $ \xi_{CB [pB ] } = 1 $. This may be used 
for this purpose with appropriate modification 
about the far-region analysis. 

As for the absorption cross-section of a minimally coupled massless scalar, 
it should become the area of the horizon in the
low energy limit \cite{DGM}. Our result about the absorption cross-sections
indicates that this kind of universality holds in a more detailed form 
for a class of black objects. One explanation is as follows.
In every case we discussed, the near-horizon wave equation reduced to the 
form like (\ref{weqRN}) and the difference among various geometries was
encoded in the term corresponding to $V_{RN}$. Since such a term 
is multiplied by the frequency as $(r_0 \omega)^2 V_{RN} $, 
the difference is becoming irrelevant and the universality appears 
as $\omega \to 0$. 

Given the CFT structure of the absorption cross-sections, one may be 
interested in its microscopic origin. 
For some parameter region of the four and five dimensional
black holes, we have the description
using $D$-branes. 
However, the corresponding microscopic theory is not known in general.
Our results include the cases which, as in \cite{MS2,CL,early}, 
are not supersymmetric 
even in the extremal limit and whose entropies near extremality do not 
have the massless ideal-gas scaling related to $p$-brane world volume
\cite{KT}.
Thus the possible underlying theory should be more general than 
the simple $p$-brane theory. The $SL(2,R)$ structure 
of the near-horizon wave equations indicates a connection to 
a microscopic theory associated with $SL(2,R)$. Probably, this is the 
string theory on $SL(2,R)$ as mentioned in \cite{CL}. 
Indeed, at extremality, the near-horizon regions of 
the four and five dimensional black holes are described 
using the $SL(2,R)$ WZW model \cite{CT}. Although a simple extension to the
non-extremal cases is impossible \cite{Russo},  a relation via 
duality transformations has been discussed  between 
the above black holes and  the BTZ black holes  
(which are locally $SL(2,R) (AdS_3)$) \cite{Hyun,SS}.
 Also, for black holes whose
near-horizon geometries are related to $SL(2,R)$, the 
entropy counting  has been done using 
a CFT associated with $SL(2,R)$ or the BTZ black holes \cite{SS,SBSS}.
These are suggestive of a connection between non-extremal black holes
and $SL(2,R)$.

The string theory on $SL(2,R)$ is not fully understood because 
its target space is a non-compact group. However, there are recent proposals
for the sensible spectrum of the strings on $SL(2,R)$ \cite{Bars,YS} 
and the BTZ black hole geometry \cite{YS}. 
It would be interesting to further investigate 
the relationship between non-extremal black holes and the string 
theory on $SL(2,R)$. 
%
%%%%%%%%%%%%%%%%%%%%%%%%%%%%%%
%\newpage
\vskip 5ex
\csectionast{Acknowledgements}
I would like to thank C.G. Callan, I.R. Klebanov, R. von Unge 
and D. Waldram for useful discussions, 
and S.P. de Alwis for correspondence.
 This work was supported in part by JSPS Postdoctral
Fellowships for Research Abroad.
%
%%%%%%%%%%%%%%%%%%%%%%%%%%%%%%%
%
% 
\newpage
%%%%%%%%%%%%%%%%%%%%%%%%%%%%%
%     references
%%%%%%%%%%%%%%%%%%%%%%%%%%%%% 
%
\def\thebibliography#1{\list
 {[\arabic{enumi}]}{\settowidth\labelwidth{[#1]}\leftmargin\labelwidth
  \advance\leftmargin\labelsep
  \usecounter{enumi}}
  \def\newblock{\hskip .11em plus .33em minus .07em}
  \sloppy\clubpenalty4000\widowpenalty4000
  \sfcode`\.=1000\relax}
 \let\endthebibliography=\endlist
\csectionast{References}
%

%
%%%%%%%%%%%%%%%%%%
\end{document}